\documentclass[review]{elsarticle}

\usepackage{hyperref}
\usepackage{url}
\usepackage{breakurl}

\journal{International Journal of Pediatric Otorhinolaryngology}

\begin{document}

\begin{frontmatter}

\title{On choking and ingestion hazards for children in the United States}

\author[mymainaddress,mysecondaryaddress]{Fr\'ed\'eric Chabolle\fnref{myfootnote}, Paul Deheuvels\corref{mycorrespondingauthor}
}
\address{Service d'ORL et de Chirurgie Cervico-Faciale, H\^opital Foch, 40 rue Worth, 92150 Suresnes, France \& L.P.S.M., Sorbonne Universit\'e, Paris, France}

\fntext[myfootnote]{Emeritus Professor}


\cortext[mycorrespondingauthor]{Corresponding author}
\ead{pr.fchabolle@gmail.com, paul.deheuvels@upmc.fr}

\address[mymainaddress]{7 avenue du Ch\^ateau, 92340 Bourg-la-Reine, France}
\address[mysecondaryaddress]{L.P.S.M., T15-25 - 2\`eme \'etage, Sorbonne Universit\'e, \\4 Place de Jussieu, 75252 Paris Cedex 05, France}

\begin{abstract}
The risks of Unintentional Suffocation injuries in the U.S. must be reconsidered in view of the existent mortality and morbidity statistics. In particular, fatal injuries due to the Sudden Unexpected Infant Death Syndrome [SUIDS] should be treated on their own and separated from this group. Because of a non-appropriate nomenclature, the risks of injuries due to Aspiration and/or Ingestion of Foreign Bodies have been overestimated in the recent decades, and should be reconsidered by a factual scrutiny of the statistical data.
\end{abstract}

\begin{keyword}
\texttt{elsarticle.cls}\sep \LaTeX\sep Elsevier \sep template
\MSC[2010] 00-01\sep  99-00
\end{keyword}

\end{frontmatter}


\section{Introduction}
In the U.S., \emph{Suffocation} has been, for long, reputed to be the leading cause of mortality and morbidity of \emph{Unintentional Injuries} in children. For example in \cite{SafeKids}, \emph{Unintentional Suffocation} is ranked as the number 1 leading cause of accidental deaths in the U.S., in 2013, for children of age \verb|<| 1 (979 deaths), as the number 3 (in a scale of decreasing gravity) for children of ages 1-4 (161 deaths), as the number 4 for children of ages 5-9 (44 deaths), and as the number 5 for children of ages 10-14 (37 deaths) (see, also, \cite{CDC-a} for counts in 2000-2006). The data displayed in \cite{SafeKids} and \cite{CDC-a} has been updated for 2017 in \cite{CDC-b}. In this last reference, in the U.S. and for 2017, \emph{Unintentional Suffocation} is ranked as the number 1 leading cause for children of age \verb|<|1 (1106 deaths), as the number 3 for children of ages 1-4 (110 deaths), as the number 4 for children of ages 5-9 (36 death), and as the number 6 for children of ages 10-14 (35 deaths). The data in
\cite{SafeKids,CDC-a,CDC-b} illustrate the stability of numbers and rankings of fatal \emph{Unintentional Suffocation} injuries in the U.S. through the last decade (see, also, \cite{Nationwide}).\vskip5pt

Among all possible causes of \emph{Unintentional Suffocation Injuries} in children, \emph{Foreign Body Aspiration} [FBA] is reputed to be a significant public health issue with presumably high mortality rate. We refer to \cite{Foltran-a,Gregori-e,Uptodate-a} for reviews and introduction to the general study of \emph{Foreign Body Injuries}. The study of FBA has given rise to considerable interest in the literature (see, e.g., \cite{Arjan,Baharloo,Baram,Berchialla,Black,Boufersaoui,Brkic,Chen,Cheng-a,Chew,
deOliveira,Eren,
Foltran-b,Foltran-c,Higo,Higuchi,Hui,Kim-b,Mallick,Mohammad,Rodriguez,Salih,Shah,
Shivakumar,Slapak,Tan,Tseng}). The annual overall inpatient cost associated with pediatric bronchial FBA is evaluated by \cite{Kim-b} to about \$12.8 million. The overall mortality of FBA injuries (with respect to all fatal and non-fatal injuries)is evaluated by \cite{Boufersaoui} to 0.29\%, by \cite{Eren} to 0.8\% and by \cite{Pasaoglu} to 0.6\%. These differences follow
from different analysis of the treatments of the FBA injuries.
\emph{Foreign Body Ingestion} [FBI] is another class of \emph{Foreign Body Injuries}. In general, Foreign Bodies can entry in the human body through various ways (aspiration, ingestion or insertion). According to \cite{Tseng}, in the U.S., in 2010, there were about 535,000 emergency department visits with foreign body-related primary diagnosis . With some exceptions, such as FBI caused by batteries, pointed objects and magnets (see, e.g., \cite{Gregori-b,Uptodate-b}), FBI generate mostly non-fatal accidents (see, e.g., \cite{Altkorn,Ambe,Blaho,Cheng-b,Gregori-b,Ikenberry,Jayachandra,Kramer,Louie,
Pasaoglu,
Singh,Velitchkov}), and the discussion in the sequel.
Some types of FBI may result in \emph{suffocation}, but such is generally not the case. We will nevertheless consider jointly FBA and FBI, to treat in the same framework all possible injuries related to \emph{Foreign Bodies}. Our interest will be centered on the age-class 0 to 19, and more specifically, to young children aged 0 to 9 who present a specific accidental behavior. A general review of FBI injuries for adults is to be found in \cite{Uptodate-b}.
The nature of FBA and FBI injuries depends largely upon the type of foreign body involved (see, e.g., \cite{Rimell}).
It may be food (see, e.g., \cite{Arjan,Altkorn,Chapin,Sidell}), non-food (see, e.g., \cite{Slapak,Sola}), or miscellaneous solid or flexible objects including balloons, coins, seeds, nuts, safety pins, pen caps, balls, marbles, toy components or small batteries (see, e.g., \cite{Ambrose,Koning,Foltran-a,Sih}). The concern about safety hazards related to choking and suffocation accidents has given rise to a number of safety alarms and epidemiological studies (see, e.g., \cite{SafeKids,CDC-a,CDC-b,WorldReport,Berchialla,Vos,Foltran-b,Gregori,Pavitt,Passali-a, Susy,Foltran-c}). Some of these investigations show clearly that the accidental risks are mostly similar between different countries, and the corresponding comparisons will not be investigated in the present paper, which will be essentially centered on the U.S. situation (see, also, \cite{Adjeso,Chinski,Chotigavanich,Higuchi,Sarafoleanu}).\vskip5pt

A close investigation of the number of casualties due to \emph{Foreign Body Aspiration or Ingestion} will show that they are largely outnumbered by other causes of \emph{Unintentional Suffocation}. In the present paper, we shall investigate the existent statistical data concerning these injuries, with emphasis on the situation of the U.S. Our findings leads us to reconsider the classification of the corresponding risks. We will show, namely, that the injuries usually grouped under the class of \emph{Unintentional Suffocation} (in the ICD-10 codes, see below) form a heterogeneous class, covering to a large extent the numerous victims of the \emph{Sudden Unexpected Infant Death Syndrome} [SUIDS] which have not much to do with the injuries induced by \emph{Foreign Bodies}. By sorting out the SUIDS from the general class of \emph{Unintentional Suffocation} injuries, we are lead to completely reconsider the rating of these risks within the class of \emph{Unintentional Injuries}.\vskip5pt

In the first place, the meaning of \emph{Unintentional Suffocation} must be made precise. It refers to the series of codes from W75 up to W84 in the 10th revision of the International Statistical Classification of Diseases [ICD-10], available at \cite{ICD-10}. The ICD coding of diseases and related health problems has been owned and copyrighted by the World Health Organization [WHO] since 1983. The WHO collects the mortality statistics corresponding to the various codes of ICD-10, which are freely available, either on the WHO Mortality Database (see, e.g., \cite{WHO-a}), or in the official databases of the different countries (for the U.S., see, e.g., \cite{WISQARS}). In either of these sites, the mortality counts are available by year, age and by ICD-10 codes. With respect to the U.S., some restrictions are imposed to the access of the counts in \cite{WISQARS}, referring to the Public Health Service Act (42 U.S.C. 242m(d)). In the present paper, we followed line by line the terms of these data use restrictions, noting that, when available, the mortality counts in \cite{WISQARS} coincide exactly with those of \cite{WHO-a}, for which no data use restriction is enforced. When such is the case, the data may be cited without any restriction. Within the U.S., the ICD-10 classification has given rise  to a national variant called the ICD-10 Clinical Modification [ICD-10-CM], which we will discuss later-on (see, e.g., \cite{ICD-10-CM}).\vskip5pt

In the ICD-10 nomenclature, \emph{Unintentional Suffocation} injuries (and in particular, \emph{Foreign Body Aspiration} injuries) are grouped within Chapter XX (\emph{External Causes of Morbidity and Mortality}, under the section \emph{Other External Causes of Accidental Injury} (codes W00-X59), and subsection \emph{Other Accidental Threats to Breathing} (codes W75-W84). We also mention the subsection \emph{Exposure to Inanimate Mechanical Forces} (codes W20-W49) of Chapter XX, which will be appropriate for \emph{Foreign Body Ingestion}. The description of these codes is as follows (refer to \cite{ICD-10,ICD-10-b}).
\begin{itemize}
\item W75.ab Accidental suffocation and strangulation in bed [ASSB]
\item W76.ab Other accidental hanging and strangulation
\item W77.ab  Threat to breathing due to cave-in, falling earth and other substances
\item W78.ab Inhalation of gastric contents
\item W79.ab Inhalation and ingestion of food causing obstruction of respiratory tract
\item W80.ab Inhalation and ingestion of other objects causing obstruction of respiratory tract
\item W81.ab Confined or trapped in a low-oxygen environment
\item W83.ab Other specified threats to breathing
\item W84.ab Unspecified threat to breathing
\end{itemize}
The ICD-10 codes covering \emph{Foreign Body Ingestion} injuries are as follows:
\begin{itemize}
\item W44.ab Foreign body entering (the body) into or through eye or natural orifice
\item W45.ab Foreign body or object entering through skin
\end{itemize}
Finally, we mention the injuries classified with Chapter XVIII \emph{Symptoms, signs and abnormal clinical and laboratory findings, not elsewhere classified}, and more specifically under the section \emph{Ill defined and unknown causes of mortality} (codes R95-R99). These are as follows.
\begin{itemize}
\item R95 Sudden infant death syndrome [SIDS]
\item R96 Other sudden death, cause unknown
\item R98 Unattended death
\item R99 Other ill-defined and unspecified causes of mortality
\end{itemize}
The values of the additional (optional) sub-codes a and b refer to the place where the injury has been inflicted and to the activity of the victim. The value of a is 0, when the injury was inflicted at home, 1 in a residential institution, 2 at school or in another institution and public administrative area, 3 in a sports and athletic area, 4 on a street or highway, 5 in a trade and service area, 6, in an industrial and construction area, 7 in a farm, 8 in an other specified place, and 9 in an unspecified place. The value of b is 0 when the victim was engaged in sports activity, 1 in leisure activity, 2 while working for income, 3 while engaged in other types of work, 4 while resting, sleeping, eating or engaging in other vital activities, 8 while engaged in other specified activities, 9 during an unspecified activity. As follows from these rules, the philosophy of ICD-10 coding is oriented towards \emph{analyzing the behavior of the victims at the time where the injuries have been inflicted}. This should allow to identify a dangerous behavior in view of future prevention of the corresponding risks. In the present paper, these sub-codes will not play any role, even though they need to be mentioned.\vskip5pt

On the opposite of ICD-10 codes, the ICD-10 CM codes are oriented towards the \emph{control of the clinical treatment of injuries, and their billable or non billable status with respect to health insurance}. Because of this, they follow different classification rules that the standard ICD-10 codes. In particular, they do not refer to the location or activity of the injury victim, as is the case with ICD-10, but rather, to the clinical description of the injury itself. In particular, they seek the explicit cause of injury when the latter is available. The ICD-10 CM codes, follow a different nomenclature and numbering than the ICD-10 codes, are as follows. First, the ICD CM code T71 is used to code \emph{Asphyxia}, namely of a severely deficient supply of oxygen that arises from abnormal breathing. It is noteworthy that \emph{Asphyxia} comprises \emph{Choking}. \emph{Asphyxia} may result in coma and death, but it is not always the case, and only a limited number of the corresponding accidents are deadly. The corresponding ICD-10-CM codes are as follows.
\begin{itemize}
\item T71 Asphyxiation
\item T71.1XXL Asphyxiation due to mechanical threat to  breathing
\item T71.11wL Asphyxiation due to smothering under pillow
\item T71.12wL Asphyxiation due to plastic bag
\item T71.13wL Asphyxiation due to being trapped in bed linens
\item T71.14wL Asphyxiation due to smothering under another person's body (in bed)
\item T71.15wL Asphyxiation due to smothering in furniture
\item T71.16wL Asphyxiation due to hanging
\item T71.19wL Asphyxiation due to mechanical threat to breathing due to other causes
\end{itemize}
The additional sub-code w takes value 1 when the cause of injury is accidental, 2 when it is intentional self-harm, 3 when it comes from an assault, 4 when the cause is undetermined. Finally X stands for blank.
Second, the ICD-10-CM Code T17 groups injuries caused by \emph{foreign bodies in respiratory tract}. These are as follows.
\begin{itemize}
\item T17 Foreign body in respiratory tract
\item T17.0XXL Foreign body in nasal sinus
\item T17.1XXL Foreign body in nostril
\item T17.2xyL Foreign body in pharynx
\item T17.3xyL Foreign body in larynx
\item T17.4xyL Foreign body in trachea
\item T17.5xyL Foreign body in bronchus
\item T17.8xyL Foreign body in other parts of respiratory tract
\item T17.9xyL Foreign body in respiratory tract, part unspecified
\end{itemize}
The sub-code x equals $0$ when the foreign body is unspecified, $1$ when it is gastric contents, $2$ when it is food, and $9$ for other foreign bodies. X stands for blank. The value of y is $0$ when the foreign body causes asphyxiation, and $8$ when it causes an other injury. Second, the ICD-CM Code T18 captures the Foreign bodies in alimentary tract. The sub-code L is a letter, taking values A for initial encounter, D for subsequent encounter, and S for sequela. We have the following family of codes.
\begin{itemize}
\item T18 Foreign body in alimentary tract
\item T18.0XXL Foreign body in mouth
\item T18.1xzL Foreign body in esophagus
\item T18.2XXL Foreign body in stomach
\item T18.3XXL Foreign body in small intestine
\item T18.4XXL Foreign body in colon
\item T18.5XXL Foreign body in anus and rectum
\item T18.8XXL Foreign body in other parts of alimentary tract
\item T18.9xyL Foreign body of alimentary tract, part unspecified
\end{itemize}
The above sub-code x equals $0$ when the foreign body is unspecified, $1$ when it is gastric contents, $2$ when it is food, and $9$ for other foreign bodies. The sub-code X stands for blank. The value of the sub-code z is $0$ when the foreign body causes compression of trachea, and $8$ when it causes an other injury. The sub-code L takes value A for initial encounter, D for subsequent encounter, and S for sequela.\vskip5pt
\noindent We must also mention the codes of the \emph{Sudden Infant Death Syndrome} [SIDS] (these are part of the general class of \emph{Sudden Unexpected Infant Death Syndrome} [SUIDS]) which has recently been included in the latest version of the ICD-10 CM coding, effective since October 1, 2018 (see, e.g., \cite{ICD-10-CM}). They are is classified under the following nomenclature.
\begin{itemize}
\item Z84.82 Sudden Infant Death Syndrome [SIDS]
\end{itemize}
 The necessity of giving the list of codes for either ICD-10 our ICD-10 CM is of some importance to illustrate the fact that both nomenclatures are essentially different, with no obvious inclusions of the ICD-10 CM codes into ICD-10 codes, and vice versa. The situation is even more complex with the ICD-10 Procedure Coding System [ICD-10-PCS] (see, e.g., \cite{ICD-10-PCS}) which is a procedure classification published by the U.S for classifying procedures performed in hospital impatient healthcare settings. The conjuction of ICD-10, ICD-10 CM and ICD-10 PCS, renders epidemiological studies delicate, to say the least. Because of this, we shall limit ourselves to the discussion of the readily on-line available official mortality statistics based upon ICD-10 codes only.

 The official mortality statistics delivered in \cite{WISQARS} by the American Centers for Disease Control and Prevention [CDC], as well as that given in \cite{WHO-a} by the World Health Organization [WHO] make a clear distinction between ICD-10 codes W75 (ASSB) and R95 (SIDS) (see above). However, it is very well possible for a medical examiner to class a cause of death into either W75 or R95 for no obvious reason. This problem is mentioned by several authors who point out on-going variations in the the classification of victims into these different ICD-10 codes (see, e.g., \cite{SIDS}). Along this line, \cite{Moon} mentions that, in the U.S., death certifiers (medical examiners and coroners) represent a diverse group with notable reporting differences. \cite{Kim} discusses the observed differentiation of victims of SUIDS in the U.S., and their repartition into the ICD-10 codes R95: sudden infant death syndrome [SIDS], R99: other ill-defined and unsuspected cause of mortality, and W75: accidental suffocation and strangulation in bed [ASSB]. They report for the two-year period 2003-2004, in the U.S., an estimated total of 7547 victims of SUIDS, out of which 2128 were listed under R99, 4408 under R95 and 931 under W75. To illustrate the reporting variability, they mention that, in 2003-2004, 45.8\% of the R99-coded deaths have been classified as “unknown”, and 48.6\% as “pending”, so that the corresponding deaths could very well have been reported under some other ICD-10 code. Likewise, among the 931 W75-coded deaths of 2003-2004, only 76.4\% were characterized as “suffocation”, so that the classification of the whole ICD-10 coded W75-deaths into SUIDS is also questionable. An update of the counts of \cite{Kim} based on the latest figures of the Centers for Disease Control and Prevention [CDC] [13] gives for 2013-2014, and children $<1$, 1674 W75-deaths (corresponding to ASSB), 3108 R95-deaths (corresponding to SIDS) and 2146 R96-98-deaths (corresponding to the other causes of SUIDS). The total cumulated count of 6928 SUIDS deaths for the two-year period 2013-2014 (close to the 7547 deaths mentioned in \cite{Kim} for 2003-2004) gives evidence that \emph{the SUIDS is, by far, the leading cause of mortality from unintentional injuries in children.}
 \vskip5pt

 As a first conclusion, we may infer from the existing literature that the precise origin of unintentional accidents related to suffocation, choking, asphyxia, foreign body aspiration or ingestion is not clear. One should base a discussion on the gravity of risks on precise counts of observed injuries, when classified into each of the existing codes. This is possible for ICD-10 codes and mortality counts, but there are some ambiguities due to SUIDS accidents where some codes clearly overlap with other codes from the "unintentional suffocation" group. The only way to clarify this situation is to make a factual discussion based on official reports. This will be made in the next section.

\section{Mortality Data}

The following displays are extracted from the data openly available on \cite{WHO-a} and \cite{WISQARS}. We start with the number of deaths classified into the ICD-10 \emph{Ill defined and unknown causes of mortality} group, comprising the codes R95-R99, among which R95 stands for the \emph{Sudden Infant Death Syndrome} (SIDS).

\begin{center}
\begin{tabular}{|c|r|r|r|r|r|r|r|r|r|r|}
  \hline
  $<$1 Year & 2007 & 2008 & 2009 & 2010 & 2011 & 2012 & 2013 & 2014 & 2015 & 2016\\ \hline
  R95 & 2453 & 2453 & 2226 & 2063 & 1910 & 1679 & 1563 & 1545 & 1568 & 1500 \\ \hline
  R96 & 0 & 0 & 0 & 0 & 0 & 0 & 0 & 0 & 0 & 0 \\
 \hline
   R98 & 5 & 2 & 5 & 3 & 5 & 3 & 2 & 2 & 2 & 3 \\
 \hline
   R99 & 1085 & 1117 & 1118 & 918 & 869 & 1061 & 1052 & 1090 & 1117 & 1247 \\
 \hline
\end{tabular}
\end{center}

\begin{center}
Table 1. SUID-related deaths $<$1 Year - ICD-10 Codes R95-R99 - 2007-2016
\end{center}

\begin{center}
\begin{tabular}{|c|r|r|r|r|r|r|r|r|r|r|}
  \hline
  1 Year & 2007 & 2008 & 2009 & 2010 & 2011 & 2012 & 2013 & 2014 & 2015 & 2016\\ \hline
  R95 & 0 & 0 & 0 & 0 & 0 & 0 & 0 & 0 & 0 & 0\\ \hline
  R96 & 27 & 36 & 32 & 30 & 43 & 31 & 33 & 28 & 26 & 38 \\
 \hline
   R98 & 3 & 1 & 3 & 0 & 0 & 1 & 1 & 1 & 0 & 3 \\
 \hline
 R99 & 93 & 106 & 105 & 101 & 114 & 111 & 114 & 89 & 120 & 112 \\
 \hline
\end{tabular}
\end{center}

\begin{center}
Table 2. SUID-related deaths 1 Year - ICD-10 Codes R95-R99 - 2007-2016
\end{center}

\begin{center}
\begin{tabular}{|c|r|r|r|r|r|r|r|r|r|r|}
  \hline
   2 Years & 2007 & 2008 & 2009 & 2010 & 2011 & 2012 & 2013 & 2014 & 2015 & 2016\\ \hline
  R95 & 0 & 0 & 0 & 0 & 0 & 0 & 0 & 0 & 0 & 0\\ \hline
  R96 & 6 & 9 & 5 & 4 & 10 & 4 & 6 & 8 & 1 & 9 \\
 \hline
   R98 & 0 & 0 & 1 & 0 & 0 & 0 & 0 & 0 & 0 & 0 \\
 \hline
 R99 & 38 & 41 & 61 & 29 & 38 & 44 & 45 & 40 & 41 & 41 \\
 \hline
\end{tabular}
\end{center}

\begin{center}
Table 3. SUID-related deaths 2 Years - ICD-10 Codes R95-R99 - 2007-2016
\end{center}

The interpretation of this data is that mortality officially attributed to SIDS (R95) is strictly limited to children of $<$1 Year. On the other side the ICD-10 code R96
(Other sudden death cause unknown) is excluded for children of $<$1 Year, but gives significative figures only for children of 1 Year. The ICD-10 code R98 (Death in circumstances where the body of the deceased was found and no cause could be discovered (found dead)) gives constantly small figures at all ages.
The remaining ICD-10 code R99 (other ill-defined and unspecified cause of mortality) is very high (of the order of 1000 deaths yearly) for children $<$1 Year, and remains important for children aged 1 and 2 (respectively of the order
of 100 and 50 deaths yearly). These figures will be compared, later-on with that relative to the ICD-10 code W75 (Accidental suffocation or strangulation in bed - ASSB). We recall from \cite{Kim} that the deaths related to SUIDS comprise primarily injuries covered by the ICD-10 codes R95, R99 and W75.\vskip5pt

We next concentrate into the mortality counts for the "unintentional suffocation group", comprising ICD-10 codes W75-W84. The next Table 4 gives data for aggregated age-groups, from 0 to 24 Years, and for Years 2010 to 2015.
\begin{center}
\begin{tabular}{|c|r|c|c|c|r|}
  \hline
  No. of deaths & $<1$ & 1-4 & 5-9 & 10-14 & 15-24 \\ \hline
  2010 & 905 & 134 & 31 & 48 & 126 \\
  2011 & 896 & 144 & 34 & 43 & 127 \\
  2012 & 965 & 138 & 34 & 45 & 124 \\
  2013 & 979 & 161 & 44 & 37 & 116 \\
  2014 & 991 & 120 & 34 & 33 & 107 \\
  2015 & 1125 & 131 & 31 & 26 & 97 \\
  \hline
\end{tabular}
\end{center}
\begin{center}
Table 4. Yearly deaths by age-groups - ICD-10 Codes W75-W84
\end{center}

It follows clearly from the data of Table 4 that mortality concentrates into the first two age classes ($<$1 and 1-4 Years old). The following data-set will discuss casualty counts for code W44, which groups all deaths due to Foreign Body Ingestion. The data gives yearly counts aggregated for children of 0-19.
\begin{center}
\begin{tabular}{|c|r|r|r|r|r|r|r|r|r|r|r|}
  \hline
  Code & 2007 & 2008 & 2009 & 2010 & 2011 & 2012 & 2013 & 2014 & 2015 & 2016 & Mean \\ \hline
  W44 & 3 & 1 & 2 & 6 & 4 & 1 & 3 & 2 & 4 & 3 & 2.7 \\ \hline
  W75 & 697 & 765 & 685 & 650 & 653 & 750 & 857 & 874 & 942 & 888 & 776.1 \\
  W76 & 100 & 86 & 92 & 88 & 80 & 80 & 69 & 74 & 64 & 60 & 79.3 \\
  W77 & 6 & 7 & 8 & 3 & 1 & 4 & 5 & 3 & 3 & 2 & 4.2 \\
  W78 & 26 & 28 & 22 & 24& 24 & 25 & 22 & 15 & 14 & 11 & 21.1 \\
  W79 & 60 & 52 & 54 & 60 & 59 & 58 & 66 & 50 & 63 & 46 & 56.8 \\
  W80 & 102 & 92 & 87 & 87 & 111 & 77 & 89 & 67 & 54 & 63 & 82.9 \\
  W81 & 10 & 3 & 2 & 4 & 1 & 2 & 0 & 3 & 1 & 0 & 2.6 \\
  W83 & 73 & 92 & 60 & 43 & 65 & 26 & 19 & 30 & 10 & 11 & 42.9 \\
  W84 & 189 & 225 & 150 & 217 & 175 & 204 & 141 & 104 & 194 & 177 & 177.6 \\ \hline
  W75-84 & 1263 & 1350 & 1160 & 1176 & 1169 & 1226 & 1268 & 1220 & 1345 & 1258 & 1243.5 \\
  \hline
\end{tabular}
\end{center}

\begin{center} Table 5. Yearly deaths by ICD-10 Codes W44 and W75-W84 - Ages 0-19
\end{center}
These displayed statistics allow us to draw three important conclusions:
\begin{itemize}
\item The number of casualties due to \emph{Foreign Body Ingestion} FBI is surprisingly small (on average, less that 3 deaths/year). This result cannot be questioned, since the ICD-10 code is non-ambiguous (Foreign body or object entering into or through eye or natural orifice. As mentioned earlier, this gives support to the idea that most FBI injuries are not deadly. Because of this, we will not discuss further FBI and limit our observation to the fact that the mortality risks of the corresponding injuries have been overestimated in the literature.
\item Deaths due to ASSB (Accidental Suffocation and Strangulation in bed), ICD-10
code W75, account (on average over the decade 2007-2016) for 62.4\% of all casualties for the whole age-group 0-19. This motivates a more detailed analysis, since obviously, ASSB should be limited to very young children.
To consider such accidental injuries for children aged (say) 19 is clearly meaningless.
\item The number of casualties due to {Foreign Body Aspiration} FBA, for the age-class 0-19, is split into three main classes of injuries. Those related to choking on food (W79 for about 79 deaths yearly), on vomit (gastric contents) (W78 for about 21 deaths yearly), and on solid (non-food or non-vomit) objects (W80 for about 83 deaths yearly). This is very far from the alarmist statements mentioned earlier
    (see, e.g., \cite{SafeKids,CDC-a,CDC-b,Nationwide}). In words, \emph{the suffocation accidents due to FBA are far from representing a leading cause of mortality in the U.S.}
\end{itemize}

The next Table 6 extends Table 5 by giving more detailed mortality counts for children aged 0 to 9, and for the aggregated ICD-10 codes W75-W84. We observe from this data that the highest number of deaths correspond to children of age $<$1
(about 980 yearly deaths). The numbers remain high for children of 1,2,3 and 4 years old (about, respectively, 73, 33, 17 and 12 yearly deaths), and it decreases with age for elder children, down to numbers of deaths in between 6 and 10 yearly.

\begin{center}
\begin{tabular}{|c|r|r|r|r|r|r|r|r|r|r|r|}
  \hline
  Age/Year & 2007 & 2008 & 2009 & 2010 & 2011 & 2012 & 2013 & 2014 & 2015 & 2016 & Mean \\ \hline
  $<$1 & 959 & 1058 & 907 & 905 & 896 & 965 & 979 & 991 & 1125 & 1023 & 980.8 \\ \hline
  1 & 92 & 86 & 62 & 72 & 71 & 67 & 87 & 58 & 70 & 62 & 72.7 \\
  2 & 38 & 29 & 34 & 29 & 38 & 40 & 32 & 27 & 36 & 28 & 33.1 \\
  3 & 10 & 20 & 20 & 13 & 23 & 18 & 20 & 15 & 14 & 20 & 17.3 \\
  4 & 9 & 10 & 9 & 20 & 12 & 13 & 22 & 20 & 11 & 8 & 12.5 \\
  5 & 7 & 12 & 6 & 2 & 6 & 9 & 5 & 9 & 7 & 5 & 6.8 \\
  6 & 13 & 2 & 4 & 6 & 7 & 6 & 5 & 7 & 8 & 7 & 6.5 \\
  7 & 4 & 7 & 4 & 5 & 7 & 6 & 12 & 5 & 5 & 6 & 6.1 \\
  8 & 9 & 9 & 4 & 10 & 5 & 3 & 6 & 3 & 5 & 8 & 6.2 \\
  9 & 9 & 11 & 8 & 8 & 9 & 10 & 16 & 10 & 6 & 9 & 9.6 \\ \hline
\end{tabular}
\end{center}

\begin{center}
Table 6. Yearly deaths 2007-2016 - 0-9 years - ICD-10 Codes W75-W84
\end{center}

Given these results, we are led to consider the impact of ICD-10 code W75 Accidental suffocation and strangulation in bed [ASSB] on the total class of unintentional suffocation accidents. The appropriate data set, extracted from \cite{WHO-a}, is displayed into the following table.

\begin{tabular}{|c|c|c|c|c|c|c|c|c|c|c|c|}
  \hline
  Age & 2005 & 2006 & 2007 & 2008 & 2009 & 2010 & 2011 & 2012 & 2013 & 2014 & Mean\\  \hline
  $<$1 & 514 & 588 & 669 & 736 & 665 & 629 & 624 & 723 & 819 & 855 & 682.2 \\  \hline
  1 & 13 & 19 & 18 & 19 & 12 & 14 & 17 & 15 & 30 & 9 & 16.6 \\
  2 & 2 & 3 & 3 & 1 & 4 & 2 & 4 & 3 & 4 & 4 & 3 \\
  3 & 2 & 1 & 2 & 1 & 2 & 1 & 1 & 3 & 1 & 0 & 1.4 \\
  4 & 2 & 0 & 0 & 1 & 2 & 0 & 1 & 1 & 1 & 0 & 0.8 \\
  5-9 & 0 & 4 & 3 & 2 & 4 & 2 & 4 & 1 & 0 & 3 & 2.3 \\
  10-14 & 2 & 2 & 2 & 2 & 0 & 0 & 1 & 3 & 0 & 1 & 1.3 \\
  15-19 & 1 & 0 & 0 & 1 & 2 & 0 & 1 & 3 & 0 & 1 & 0.9 \\
  \hline
\end{tabular}

\begin{center}
Table 7. Yearly deaths 2005-2014 - 0-19 years - ICD-10 Code W75
\end{center}

The counts in Table 7 leave open some classification questions. Obviously,
the mortality data of ASSB pertaining to children of age $<$1 Year can be included in the class of SUIDS accidents. This corresponds to a 10-year mean of casualties as high as 96.29\% of all deaths within the age-class 0-19. The question is not at all obvious from
the data corresponding to higher ages. In particular, the classification of ASSB injuries pertaining to children aged 15-19 into SUIDS is very much questionable. The necessity of a more precise classification is illustrated by these examples, which, fortunately concern very few accidents. This data gives the following conclusion.

\begin{itemize}
\item The mortality data pertaining to the ICD-10 code W75 (Accidental suffocation and strangulation in bed [ASSB]) cannot be aggregated with unintentional suffocation data pertaining to other causes. It should be incorporated into the SUIDS group, with some specific treatment allocated to the injuries inflicted to children older than 3.
\end{itemize}
We now rejoin our discussion concerning \emph{Foreign Body Aspiration}. As mentioned earlier, this concerns the ICD-10 codes W78, W79 and W80. The following tables give the corresponding mortality counts for children aged 0 to 9.

\begin{center}
\begin{tabular}{|c|r|r|r|r|r|r|r|r|r|r|r|}
  \hline
  Age/Year & 2007 & 2008 & 2009 & 2010 & 2011 & 2012 & 2013 & 2014 & 2015 & 2016 & Mean \\ \hline
  $<$1 & 15 & 19 & 9 & 12 & 14 & 12 & 11 & 8 & 8 & 4 & 11.4 \\ \hline
  1 & 4 & 5 & 4 & 4 & 3 & 4 & 4 & 1 & 3 & 2 & 3.4 \\
  2 & 1 & 0 & 2 & 4 & 0 & 2 & 3 & 1 & 0 & 0 & 1.3 \\
  3 & 0 & 0 & 2 & 0 & 0 & 0 & 0 & 1 & 0 & 0 & 0.3 \\ \hline
  4 & 0 & 0 & 0 & 0 & 0 & 0 & 0 & 2 & 0 & 1 & 0.3 \\ \hline
  5 & 0 & 0 & 0 & 0 & 0 & 0 & 0 & 1 & 0 & 1 & 0.2 \\
  6 & 1 & 0 & 0 & 0 & 1 & 1 & 0 & 0 & 0 & 1 & 0.4 \\
  7 & 0 & 0 & 0 & 0 & 1 & 0 & 0 & 0 & 0 & 0 & 0.1 \\
  8 & 0 & 1 & 0 & 0 & 0 & 0 & 1 & 0 & 0 & 1 & 0.3 \\
  9 & 0 & 0 & 0 & 1 & 0 & 1 & 0 & 0 & 0 & 0 & 0.2 \\ \hline
\end{tabular}
\end{center}

\begin{center} Table 8. Yearly deaths 2007-2016 - 0-9 Years - ICD-10 Code W78 (choking on vomit)
\end{center}

\begin{center}
\begin{tabular}{|c|r|r|r|r|r|r|r|r|r|r|r|}
  \hline
  Age/Year & 2007 & 2008 & 2009 & 2010 & 2011 & 2012 & 2013 & 2014 & 2015 & 2016 & Mean \\ \hline
  $<$1 & 13 & 10 & 9 & 12 & 9 & 11 & 9 & 10 & 14 & 7 & 10.4 \\ \hline
  1 & 19 & 16 & 13 & 14 & 15 & 16 & 22 & 13 & 20 & 11 & 15.9 \\
  2 & 17 & 10 & 13 & 8 & 13 & 18 & 13 & 9 & 7 & 11 & 11.9 \\
  3 & 2 & 5 & 4 & 5 & 3 & 4 & 5 & 5 & 6 & 5 & 4.4 \\ \hline
  4 & 3 & 1 & 1 & 7 & 4 & 1 & 9 & 2 & 4 & 0 & 3.2 \\ \hline
  5 & 1 & 0 & 2 & 0 & 1 & 1 & 0 & 2 & 1 & 1 & 0.9 \\
  6 & 1 & 0 & 0 & 2 & 1 & 0 & 0 & 1 & 3 & 2 & 1 \\
  7 & 0 & 2 & 0 & 1 & 3 & 0 & 0 & 1 & 1 & 0 & 0.8 \\
  8 & 0 & 1 & 0 & 1 & 0 & 0 & 0 & 0 & 0 & 0 & 0.2 \\
  9 & 0 & 0 & 0 & 1 & 2 & 1 & 1 & 1 & 0 & 2 & 0.8 \\ \hline
\end{tabular}
\end{center}

\begin{center}
Table 9. Yearly deaths 2007-2016 - 0-9 years - ICD-10 Code W79 (choking on food)
\end{center}

\begin{center}
\begin{tabular}{|c|r|r|r|r|r|r|r|r|r|r|r|}
  \hline
  Age/Year & 2007 & 2008 & 2009 & 2010 & 2011 & 2012 & 2013 & 2014 & 2015 & 2016 & Mean \\ \hline
  $<$1 & 42 & 33 & 36 & 34 & 48 & 22 & 25 & 22 & 19 & 22 & 30.3 \\ \hline
  1 & 22 & 15 & 14 & 17 & 14 & 12 & 16 & 12 & 9 & 12 & 14.3 \\
  2 & 6 & 6 & 5 & 10 & 13 & 10 & 5 & 4 & 7 & 3 & 6.9 \\
  3 & 2 & 4 & 3 & 1 & 11 & 2 & 5 & 2 & 4 & 4 & 3.8 \\ \hline
  4 & 1 & 3 & 3 & 6 & 5 & 7 & 6 & 4 & 3 & 2 & 4 \\ \hline
  5 & 2 & 4 & 2 & 1 & 3 & 3 & 2 & 0 & 0 & 1 & 1.8 \\
  6 & 5 & 2 & 1 & 0 & 1 & 2 & 2 & 3 & 0 & 3 & 1.9 \\
  7 & 0 & 3 & 1 & 0 & 0 & 1 & 5 & 0 & 2 & 1 & 1.3 \\
  8 & 2 & 1 & 2 & 1 & 0 & 2 & 2 & 1 & 2 & 1 & 1.4 \\
  9 & 5 & 1 & 1 & 1 & 1 & 1 & 0 & 1 & 1 & 0 & 1.2 \\ \hline
\end{tabular}
\end{center}

\begin{center}
Table 10. Yearly deaths 2007-2016 - 0-9 years - ICD-10 Code W80 (choking on non-food items)
\end{center}

The conclusion from the data in Tables 8-10 is as follows.

\begin{itemize}
\item Foreign body aspiration [FBA] injuries can be split into three main classes corresponding to the different types of bodies involved.
\item The inhalation of gastric contents (vomit) is covered by ICD-10 code W78. The risk is relatively high for children $<$1 Year (about 11 deaths/yearly). It diminishes by half for children 1-3 (about 5 deaths/yearly). For elder children ($\geq$4) it remains at a very low level (about 2 deaths yearly).
\item The inhalation of food is covered by ICD-10 code W78. The risk is relatively high for children of age 0-4 with a peak for 1 year-old children (about 46 deaths yearly). For children of 5 and above, the risk is low (about 5 deaths yearly).
\item The inhalation of non-food objects is covered by ICD-10 code W80. The risk is relatively high for children of age 0-3 (about 55 deaths yearly).
    It diminishes for children aged 4-9 to an overall mean of about 12 deaths yearly.
\item These numbers show that the risk of foreign body aspiration (and ingestion) has been very much overestimated in the literature. The error originates from the aggregation of this risk with a series of other related risks not specifically related to FBA and FBI.
\end{itemize}

\section{Fatal versus Nonfatal Data}

The ratio between fatal and nonfatal data is commonly used to provide estimates
of nonfatal injuries given the exact counts of fatal injuries. Some of this data is available in \cite{CDC-c,CDC-d}. Unfortunately, the openly available counts are given for grouped data, namely for the \emph{Unintentional Suffocation} injuries, corresponding to the ICD-10 codes W75-W84. In 2017, for a U.S. population of 325,719,178 inhabitants, one so obtains 6,946 deaths, for 57,137 nonfatal injuries. This corresponds to data based on all ages. Naturally for the children age-group, these numbers must be modified (see below). The ratio (nonfatal)/(fatal) is therefore equal to (about) 8.22. Or, likewise, the ratio (fatal)/(nonfatal) is equal to (about) 12.15\%.
This is far from the estimations of \cite{Boufersaoui,Eren,Pasaoglu} who are close to multiplying by 10 the official  figures of nonfatal unintentional suffocation given
in \cite{CDC-d}. We note that the latter evaluations correspond to a collection of data made by the National Electronic Injury surveillance System [NEISS] to
monitor nonfatal injuries treated in U.S. hospital emergency departments. Obviously, the ratio (nonfatal)/(fatal) must vary with respect of the different ICD-10 codes. For example, those who refer to SUIDS will be equal to 0, all accidents being, by definition, fatal. For this reason, we can only offer a thumb-rule to evaluate the yearly number of FBA nonfatal injuries. Below are given more detailed displays.\vskip10pt

\begin{tabular}{|c|c|c|c|c|c|c|c|c|c|}
  \hline
    Age & 2007 & 2008 & 2009 & 2010 & 2011 & 2012 & 2013 & 2014 & 2015 \\  \hline
  00-04 & 1108 & 1203 & 1032 & 1039 & 1040 & 1103 & 1140 & 1111 & 1256  \\
  05-09 & 42 & 41 & 26 & 31 & 34 & 34 & 44 & 34 & 31  \\
  10-14 & 60 & 50 & 41 & 48 & 43 & 45 & 37 & 33 & 26   \\
  15-19 & 53 & 56 & 61 & 58 & 52 & 44 & 47 & 42 & 32 \\
  Total & 1263 & 1350 & 1160 & 1176 & 1169 & 1226 & 1268 & 1220 & 1345  \\
  \hline
\end{tabular}

\begin{center}
Table 11. Yearly deaths 2007-2015 - 0-19 years - ICD-10 Codes W75-W84
\end{center}
\vskip10pt

\begin{tabular}{|c|c|c|c|c|c|c|c|c|c|}
  \hline
  Age &  2007 & 2008 & 2009 & 2010 & 2011 & 2012 & 2013 & 2014 & 2015 \\  \hline
  00-04  & 12236 & 10543 & 11820 & 13804 & 16429 & 17603 & 17679 & 1256 & 1141  \\
  05-09  & 41 & 26 & 31 & 34 & 34 & 44 & 34 & 31 & 35  \\
  10-14  & 50 & 41 & 48 & 43 & 45 & 37 & 33 & 26 & 39  \\
  15-19  & 56 & 61 & 58 & 52 & 44 & 47 & 42 & 32 & 43 \\
  Total  & 1350 & 1160 & 1176 & 1169 & 1226 & 1268 & 1220 & 1345 & 1258 \\
  \hline
\end{tabular}

\begin{center}
Table 12. Nonfatal Injuries 2007-2015 - 0-19 years - ICD-10 Codes W75-W84
\end{center}

By considering the age-groups in Tables 11-12, we conclude that
the ratio (nonfatal injuries)/(fatal injuries) is of the order of 50 for the age-group
5-19.


\section{Conclusion}
The accidental data provided in the present paper illustrates the urgent need of a precise nomenclature of the types of injuries to be considered. On the whole, the only precise available statistics are the mortality counts pertaining to each ICD code. The number of non-fatal injuries can then be deduced from these counts given appropriate estimations of the ratio (Non-fatal injuries)/(Fatal injuries). To achieve a proper epidemiological analysis, it is essential for the ICD nomenclature in force to avoid any possible ambiguity, such as that induced by the possibility of reporting an accident under different ICD codes. The case of the Sudden Unexpected Infant Death Syndrome [SUIDS] is a perfect example of the problem. By reporting a number of SUIDS (but not all) as Accidental Suffocation and Strangulation in Bed [ASSB], then, by aggregating the corresponding ICD-10 code W75 into the Unintentional Suffocation group W75-W84, the official organisms responsible for accidental statistics have propagated a distorted view of the real situation of accidental risks. Because of this, the risks of Foreign Body Aspiration and Ingestion [FBA-FBI] has been largely overestimated, giving rise to unappropriate alarms. Clearly, the official organisms collecting accidental data cannot be considered as responsible for this major error, but the conclusion is that one should remain cautious in the future of possible false alarms due to the nomenclature of accidents. We hope sincerely that the newer versions of ICD codes will bring improvements to the unfortunate oversights brought by ICD-9 and ICD-10 coding systems. The simple comparison of ICD-10 codes with ICD-10 CM and ICD-10 PCS codes shows that these three different coding systems are not fully coherent. Some efforts should be made to correct this situation, otherwise they should remain mutually incompatible.

\section*{References}

\bibliography{mybibfile2}

\begin{thebibliography}{10}
\expandafter\ifx\csname url\endcsname\relax
  \def\url#1{\texttt{#1}}\fi
\expandafter\ifx\csname urlprefix\endcsname\relax\def\urlprefix{URL }\fi
\expandafter\ifx\csname href\endcsname\relax
  \def\href#1#2{#2} \def\path#1{#1}\fi

\bibitem{SafeKids}
{Safe Kids Worldwide}, Overview of global childhood injury morbidity and
  mortality fact sheet 2014, \url{https://www.safekids.org/fact-sheet/
  overview-global-childhood}\url{-injury-morbidity-and-mortality-fact}
  \url{-sheet-2014-pdf/}, Last accessed on 2019-05-01 (2015).

\bibitem{CDC-a}
N.~Borse, J.~Gilchrist, A.~Dellinger, R.~Rudd, M.~Ballesteros, D.~Sleet, {CDC}
  {C}hildhood {I}njury {R}eport: {Patterns of Unintentional Injuries among 0-19
  Years Olds in the United States, 2000-2006}, atlanta (GA): {Centers for
  Disease Control and Prevention, National Center for Injury Prevention and
  Control} \url{https://www.cdc.gov/safechild/pdf/cdc-childhoodinjury.pdf},
  Last accessed on 2019-05-01 (2008).

\bibitem{CDC-b}
{{CDC - Centers for Disease Control and Prevention} {WISQARS - Web-based Injury
  Statistics Query and Reporting System} {National Center for Injury Prevention
  and Control}}, {{10 Leading Causes of Injury Deaths by Age Group Highlighting
  Unintentional Injury Deaths, United States - 2017}},
  \url{https://www.cdc.gov/injury/wisqars/LeadingCauses.html}, Last accessed on
  2019-05-01 ({2017}).

\bibitem{Nationwide}
{Nationwide {C}hildren's {H}ospital}, Choking is a leading cause of injury and
  death among children,
  \url{www.sciencedaily.com/releases/2010/02/100226212559.htm}, Last accessed
  on 2019-05-01 (28 {F}ebruary 2010).

\bibitem{Foltran-a}
F.~Foltran, D.~Gregori, D.~Pass{\`a}li, L.~Bellussi, G.~Caruso, F.~Pass{\`a}li,
  G.~Pass{\`a}li, the {ESFBI}~{S}tudy {G}roup, Toys in the upper aerodigestive
  tract: {E}vidence on their risk as emerging from the {ESFBI} study, Auris
  Nasus Larynx 38 (2011) 612--617.

\bibitem{Gregori-e}
D.~Gregori, The {S}usy {S}afe {P}roject: {A} web-based registery of foreign
  body injuries in children, International Journal of Pediatric
  Otorhinolaryngology 70 (2006) 1663--1664.

\bibitem{Uptodate-a}
F.~Ruiz, G.~Mallory, S.~Torrey, A.~Hoppin, {Airway foreign bodies in children},
  {{UpToDate}} (2018) {1--15,
  }\url{https://www.uptodate.com/contents/}\url{airway-foreign-bodies-in-children},
  Last accessed on 2019-05-01.

\bibitem{Arjan}
B.~Arjan, A.~Yusof, A.~Millar, T.~S. S.~W. Group, Food foreign body injuries,
  International Journal of Pediatric Otorhinolaryngology 76S (2012) S20--S25.

\bibitem{Baharloo}
F.~Baharloo, F.~Veyckemans, C.~Francis, M.-P. Biettlot, D.~Rodenstein,
  Tracheobronchial foreign bodies - presentation and management in children and
  adults, CHEST 115~(5) (1999) 1357--1362.

\bibitem{Baram}
A.~Baram, H.~Sherzad, S.~Saeed, F.~Kakamad, A.~Hamawandi, Tracheobronchial
  foreign bodies in children: the role of emergency rigid bronchoscopy, Global
  Pediatric Health 4 (2017) 1--6.
\newblock \href {http://dx.doi.org/10.1177/2333794X17743663}
  {\path{doi:10.1177/2333794X17743663}}.

\bibitem{Berchialla}
P.~Berchialla, L.~Bellussi, A.~Castella, S.~Snidero, D.~Passali, D.~Gregori,
  Modeling the risk: Innovative approaches to understand and quantify the risk
  of severe {FB} injury, International Journal of Pediatric Otorhinolaryngology
  76S (2012) S33--S38.

\bibitem{Black}
R.~Black, D.~Johnson, M.~Matlack, Bronchoscopic removal of aspirated foreign
  bodies in children, Journal of Pediatric Surgery 29~(5) (1994) 682--684.

\bibitem{Boufersaoui}
A.~Boufersaoui, L.~Smati, K.~Benhalla, R.~Boukari, S.~Smail, K.~Anik,
  R.~Aouameur, H.~Chaouche, M.~Baghriche, Foreign body aspiration in children:
  Experience from 2624 patients, International Journal of Pediatric
  Otorhinolaryngology 77 (2013) 1683--1688.

\bibitem{Brkic}
F.~Brki{\'c}, {\u{S}}.~Umihani{\'c}, Tracheobronchial foreign bodies in
  children - experience at {ORL} clinic {T}uzla, 1995--2004, International
  Journal of Pediatric Otorhinolaryngology 71 (2007) 909--915.

\bibitem{Chen}
X.~Chen, C.~Zhang, Foreign body aspiration in children: {F}ocus on the impact
  of delayed treatment, International Journal of Pediatric Otorhinolaryngology
  71 (2017) 111--115.

\bibitem{Cheng-a}
J.~Cheng, B.~Liu, A.~Farjat, J.~Routh, The public health resource utilization
  impact of airway foreign bodies in children, International Journal of
  Pediatric Otorhinolaryngology 96 (2017) 68--71.

\bibitem{Chew}
H.~Chew, H.~Tan, Airway foreign body in children, International Journal of
  Clinical Medicine 3 (2012) 655--660.

\bibitem{deOliveira}
C.~de~Oliveira, J.~de~Almeida, E.~Troster, F.~Vaz, Complications of
  tracheobronchial foreign body aspiration in children: {R}eport of 5 cases and
  review of the literature, Revista do Hospital das Cl{\'{\i}}nicas
  Universidade de S{\~{a}}o Paulo Hospital das Cl{\'{\i}}nicas 57~(3) (2002)
  108--111.

\bibitem{Eren}
S.~Eren, A.~Balci, B.~Dikici, M.~Doblan, M.~Eren, Foreign body aspiration in
  children: {E}xperience of 1160 cases, Annals of Tropical Paediatrics 23~(1)
  (2003) 31--37.

\bibitem{Foltran-b}
F.~Foltran, S.~Ballali, F.~Passali, E.~Kern, B.~Morra, G.~Passali,
  P.~Berchialla, M.~Lauriello, D.~Gregori, Foreign bodies in the airways: {A}
  meta-analysis of published papers, International Journal of Pediatric
  Otorhinolaryngology 76S (2012) S12--S19.

\bibitem{Foltran-c}
F.~Foltran, P.~Berchialla, D.~Gregori, A.~Pitk{\"a}ranta, I.~Slapak,
  J.~Jakub{\'{\i}}kov{\'a}, L.~Bellussi, D.~Passali, Stationery injuries in the
  upper aerodigestive system: {R}esults from the {S}usy {S}afe {P}roject,
  International Journal of Pediatric Otorhinolaryngology 76S (2012) S67--S72.

\bibitem{Higo}
R.~Higo, Y.~Matsumoto, K.~Ichimura, K.~Kaga, Foreign bodies in the
  aerodigestive tract in pediatric patients, Auris Nasus Larynx 30 (2003)
  397--401.

\bibitem{Higuchi}
O.~Higuchi, Y.~Adachi, T.~Ichimaru, M.~Asai, K.~Kawasaki, Foreign body
  aspiration in children: {A} nationwide survey in {J}apan, International
  Journal of Pediatric Otorhinolaryngology 73 (2009) 659--661.

\bibitem{Hui}
H.~Hui, L.~Na, C.~Zhijun, Z.~Fugao, S.~Yan, Z.~Niankai, C.~Jinggjing,
  Therapeutic experience from 1428 patients with pediatric tracheobronchial
  foreign body, Journal of Pediatric Surgery 43 (2008) 718--721.

\bibitem{Kim-b}
I.~Kim, N.~Shapiro, N.~Bhattacharyya, The national cost burden of bronchial
  foreign body aspiration in children, Laryngoscope 125 (2015) 1221--1224.

\bibitem{Mallick}
M.~Mallick, Tracheobronchial foreign body aspiration in children: {A}
  continuing diagnostic challenge, African Journal of Paediatric Surgery
  11~(3).

\bibitem{Mohammad}
M.~Mohammad, M.~Saleem, M.~Mahseeri, I.~Alabdallat, A.~Alomari, A.~Za'atreh,
  I.~Qudaisat, A.~Shufidat, M.~Alzoubi, Foreign body aspiration in children:
  {A} study of children who lived or died following aspiration, International
  Journal of Pediatric Otorhinolaryngology 98 (2017) 29--31.

\bibitem{Rodriguez}
H.~Rodriguez, G.~Passali, D.~Gregori, Management of foreign bodies in the
  airway and oesophagus, International Journal of Pediatric Otorhinolaryngology
  76S (2012) S84--S91.

\bibitem{Salih}
A.~Salih, M.~Alfaki, D.~Alam-Elhuda, Airway foreign bodies: {A} critical review
  for a common pediatric emergency, World Journal of Emergency Medicine 7~(1)
  (2016) 5--12.

\bibitem{Shah}
R.~Shah, A.~Patel, L.~Lander, S.~Choi, Management of foreign bodies obstructing
  the airway in children, Archives of Otorhinolaryngology - Head and Neck
  Surgery 136~(4) (2010) 373--379.

\bibitem{Shivakumar}
A.~Shivakumar, A.~Naik, K.~Prashanth, K.~Shetty, D.~Praveen, Tracheobronchial
  foreign bodies, Indian Journal of Pediatrics 70~(10) (2003) 793--797.

\bibitem{Slapak}
I.~Slapak, F.~Passali, A.~Gulati, {the Susy Safe Working Group}, Non food
  foreign body injuries, International Journal of Pediatric Otorhinolaryngology
  76S (2012) S26--S32.

\bibitem{Tan}
H.~Tan, K.~Brown, T.~{McGill}, M.~Kenna, D.~Lund, G.~Healy, Airway foreign
  bodies {(FB)}: {A} 10-year review, International Journal of Pediatric
  Otorhinolaryngology 56 (2000) 91--99.

\bibitem{Tseng}
H.-J. Tseng, T.~Hanna, W.~Shuaib, F.~Khosa, K.~Linnau, Imaging foreign bodies:
  {I}ngested, aspirated and inserted, Annals of Emergency Medicine 66~(6)
  (2015) 570--582.

\bibitem{Pasaoglu}
I.~Pa{\c{s}}ao{\v{g}}lu, R.~Do{\v{g}}an, M.~Demircin, Bronchoscopic removal of
  foreign bodies in children: {R}etrospective analysis of 822 cases, Thoracic
  Cardiovascular Surgery 39~(2) (1991) 95--98.

\bibitem{Gregori-b}
D.~Gregori, B.~Morra, A.~Gulati, Magnetic {FB} injuries: {A}n old yet
  unresolved hazard, International Journal of Pediatric Otorhinolaryngology 76S
  (2012) S42--S48.

\bibitem{Uptodate-b}
G.~Triadafilopoulos, J.~Saltzman, S.~Grover, {Ingested foreign bodies and food
  impactions in adults}, {{UpToDate}} (2018) {1--15,
  }\url{https://www.uptodate.com/contents/}\url{ingested-foreign-bodies-and-food-
  impactions}\url{-in-adults}, Last accessed on 2019-05-01.

\bibitem{Altkorn}
R.~Altkorn, X.~Chen, S.~Milkovich, D.~Stool, G.~Rider, C.~Bailey, A.~Haas,
  K.~Riding, S.~Pransky, J.~Reilly, Fatal and non-fatal food injuries among
  children (aged 0--14 years), International Journal of Pediatric
  Otorhinolaryngology 72 (2008) 1041--1046.

\bibitem{Ambe}
P.~Ambe, S.~Weber, M.~Schauer, W.~Knoefel, Swallowed foreign bodies in adults,
  Deutsches Arzteblatt International 109~(50) (2012) 2869--875.

\bibitem{Blaho}
K.~Blaho, K.~Merigian, S.~Winbery, L.~Park, M.~Cockrell, Foreign body
  ingestions in the emergency department: case reports and review of treatment,
  Journal of Emergency Medicine 16~(1) (1998) 21--26.

\bibitem{Cheng-b}
W.~Cheng, P.~Tam, Foreign body ingestion in children: {E}xperience with 1,265
  cases, Journal of Pediatric Surgery 34~(10) (1999) 1472--1476.

\bibitem{Ikenberry}
S.~Ikenberry, T.~Jue, V.~Appelaneni, S.~Benerjee, T.~Ben-Menachem, A.~Decker,
  R.~Fanelli, L.~Fisher, N.~Fukami, M.~Harrison, R.~Jain, K.~Khan, M.~Krinsky,
  J.~Maple, R.~Sharaf, L.~Strohmeyer, J.~Dominitz, {{A.G.S.E.} {S}tandards of
  {P}ractice {C}ommittee}, Guideline - {M}anagement of ingested foreign bodies
  and food impactations, Gastrointestinal Endoscopy 73~(6) (2011) 1085--1091.
\newblock \href {http://dx.doi.org/10.1016/j.gie.2010.11.010}
  {\path{doi:10.1016/j.gie.2010.11.010}}.

\bibitem{Jayachandra}
S.~Jayachandra, G.~Eslick, A systematic review of paediatric foreign body
  ingestion: {P}resentation, complication, and management, International
  Journal of Pediatric Otorhinolaryngology 77 (2013) 311--317.

\bibitem{Kramer}
T.~Kramer, K.~Reiding, l.J. Salkeld, Tracheobronchial and esophagus foreign
  bodies in the pediatric population, Journal of Otorhinolaryngology 15~(6)
  (1986) 355--358.

\bibitem{Louie}
M.~Louie, S.~Bradin, Foreign body ingestion and aspiration, Pediatric Review
  30~(8) (2009) 295--301.

\bibitem{Singh}
G.~Singh, S.~Sharma, S.~Khurade, S.~Gooptu, Ingested foreign bodies in
  children: {A} report of two cases, Journal of Family Medicine and Primary
  Care 3~(4) (2014) 452--455.
\newblock \href {http://dx.doi.org/10.4103/2249-4863.148148}
  {\path{doi:10.4103/2249-4863.148148}}.

\bibitem{Velitchkov}
N.~Velitchkov, G.~Grigorov, J.~Losanoff, K.~Kjossev, Ingested foreign bodies of
  the gastrointestinal tract: {R}etrospective analysis of 542 cases, World
  Journal of Surgery 20 (1996) 1001--1005.

\bibitem{Rimell}
F.~Rimell, A.~Thorme, S.~Stool, J.~Reilly, G.~Rider, D.~Stool, C.~Wilson,
  Characteristics of objects that cause choking in children, JAMA 274 (1995)
  1763--1766.

\bibitem{Chapin}
M.~Chapin, L.~Rochette, J.~Annest, T.~Haileyesus, K.~Conner, G.~Smith, Nonfatal
  choking on food among children 14 years of younger in the {U}nited {S}tates,
  2001--2009, Pediatrics 132~(2) (2013) 275--281.

\bibitem{Sidell}
D.~Sidell, I.~Kim, T.~Coker, C.~Moreno, N.~Shapiro, Food choking hazards in
  children, International Journal of Pediatric Otorhinolaryngology 77 (2013)
  1940--1946.

\bibitem{Sola}
R.~{Sola Jr}, E.~Rosenfeld, Y.~Yu, S.~S. Peter, S.~Shah, Magnet foreign body
  ingestion: rare occurrence but big consequences, Journal of Pediatric Surgery
  53~(9) (2018) 1815--1819.
\newblock \href {http://dx.doi.org/10.1016/j.pedsurg.2017.08.013}
  {\path{doi:10.1016/j.pedsurg.2017.08.013}}.

\bibitem{Ambrose}
S.~Ambrose, N.~Raol, Pediatric airway foreign body, Operative Techniques in
  Otolaryngology 28 (2017) 265--269.

\bibitem{Koning}
T.~de~Koning, F.~Foltran, D.~Gregori, Fostering design for avoiding small parts
  in commonly used objects, International Journal of Pediatric
  Otorhinolaryngology 76S (2012) S57--S60.

\bibitem{Sih}
T.~Sih, C.~Bunnag, S.~Ballali, M.~Lauriello, L.~Bellussi, Nuts and seed: {A}
  natural yet dangerous foreign body, International Journal of Pediatric
  Otorhinolaryngology 76S (2012) S49--S52.

\bibitem{WorldReport}
M.~Peden, K.~Oyegbite, J.~Ozanne-Smith, A.~Hyder, C.~Branche, A.~Rahman,
  F.~Rivara, K.~Bartolomeos, {World report on child injury prevention}, {{WHO
  Library Cataloguing-in-Publication Data - World Health Organization}}, 2008.

\bibitem{Vos}
T.~Vos, the Global Burden~of Disease Pediatrics~Collaboration, Global and
  national burden of diseases and injuries among children and adolescents
  between 1990 and 2013 {F}indings from the global burden of disease 2013
  study, JAMA Pediatrics 170~(3) (2016) 267--287.
\newblock \href {http://dx.doi.org/10.1001/jamapediatrics.2015.4276}
  {\path{doi:10.1001/jamapediatrics.2015.4276}}.

\bibitem{Gregori}
D.~Gregori, D.~Passali, Foreign bodies injuries: {A} strong unique pathway
  linking {ORL} and public health, International Journal of Pediatric
  Otorhinolaryngology 76S (2012) S1.

\bibitem{Pavitt}
M.~Pavitt, J.~Nevett, L.~Swanton, M.~Hind, M.~Polkey, M.~Green, N.~Hopkinson,
  London ambulance source data on choking incidence for the calendar year 2016:
  an observational study, {BMJ} Open Respiratory Research 138 (2017) 1--4.
\newblock \href {http://dx.doi.org/10.1136/bmjresp-2017-000215}
  {\path{doi:10.1136/bmjresp-2017-000215}}.

\bibitem{Passali-a}
D.~Passali, Foreign body injuries: The urgent need for updating the field,
  International Journal of Pediatric Otorhinolaryngology 76S  S2.

\bibitem{Susy}
{The Susy Safe Working Group}, The {S}usy {S}afe project overview after the
  first four years of activity, International Journal of Pediatric
  Otorhinolaryngology 76S (2012) S3--S11.

\bibitem{Adjeso}
T.~Adjeso, M.~Damah, J.~Murphy, T.~Anyomih, Foreign body aspiration in northern
  {G}hana: A review of pediatric patients, International Journal of
  Otolaryngology 2017~(1--4).
\newblock \href {http://dx.doi.org/10.1155/2017/1478795}
  {\path{doi:10.1155/2017/1478795}}.

\bibitem{Chinski}
A.~Chinski, F.~Foltran, D.~Gregori, S.~Ballali, D.~Passali, L.~Bellussi,
  Foreign bodies in children: {A} comparison between {A}rgentina and {E}urope,
  International Journal of Pediatric Otorhinolaryngology 76S (2012) S76--S79.

\bibitem{Chotigavanich}
C.~Chotigavanich, S.~Ballali, F.~Foltran, D.~Passali, L.~Bellussi, D.~Gregori,
  the {ESFBI}~{S}tudy {G}roup, Foreign bodies injuries in children: {A}nalysis
  of {T}hailand data, International Journal of Pediatric Otorhinolaryngology
  76S (2012) S80--S83.

\bibitem{Sarafoleanu}
C.~Sarafoleanu, S.~Ballali, D.~Gregori, L.~Bellussi, D.~Passali, Retrospective
  study on {R}omanian foreign body injuries in children, International Journal
  of Pediatric Otorhinolaryngology 76S (2012) S73--S75.

\bibitem{ICD-10}
{{World Health Organization}}, {{International Statistical Classification of
  Diseases and Related Health Problems - 10th Revision - Volume 2 Instruction
  Manual, Edition 2010}}, \url{https://icd.who.int/browse10/Content/statichtml/
  ICD10Volume2_en_2010.pdf}, Last accessed on 2019-05-01 ({2010}).

\bibitem{WHO-a}
{{World Health Organization}}, {{WHO Mortality Database}},
  \url{http://apps.who.int/healthinfo/statistics/mortality/
  whodpms/help/help.htm}, Last accessed on 2019-05-01 (2019).

\bibitem{WISQARS}
{{Centers for Disease Control and Prevention}}, {{Fatal Injury Reports,
  National, Regional and State, 1981-2017}},
  \url{https://webappa.cdc.gov/sasweb/ncipc/mortrate.html}, Lasr accessed on
  2019-05-01 (2019).

\bibitem{ICD-10-CM}
{{CDC - Centers for Disease Control and Prevention} {National Center for Health
  Statistics}}, {{International Classification of Diseases, Tenth Revision,
  Clinical Modification (ICD-10-CM)}},
  \url{https://www.cdc.gov/nchs/icd/icd10cm.htm#FY 2019 release of ICD-10-CM},
  Last accessed on 2019-05-01 ({2019}).

\bibitem{ICD-10-b}
{{World Health Organization}}, {{International Statistical Classification of
  Diseases and Related Health Problems 10th Revision - ICD-10 Version:2016}},
  \url{https://icd.who.int/browse10/2016/en}, Last accessed on 2019-05-01
  (2016).

\bibitem{ICD-10-PCS}
{{CMS.gov Centers for Medicare and Medicaid Services}}, {{The 2019 ICD-10
  Procedure Coding System (ICD-10-PCS)}},
  \url{https://www.cms.gov/Medicare/Coding/ICD10/2019-ICD-10-PCS.html}, Last
  accessed on 2019-05-01 ({2019}).

\bibitem{SIDS}
{{Task Force on Sudden Infant Death Syndrome}}, The changing concept of sudden
  infant death syndrome: {D}iagnostic coding shift, controversies regarding the
  sleeping environment and new variables to consider in reducing risks,
  Pediatrics 116 (2005) 1245--1255.

\bibitem{Moon}
{R.Y. Moon and {AAP Task Force on Sudden Infant Death Syndrome}}, {SIDS} and
  other sleep-related infant deaths: evidence base for 2016 – updated
  recommendations for a safe infant sleeping environment, Pediatrics 138 (2016)
  1--34.

\bibitem{Kim}
S.~Kim, C.~Shapiro-Mendoza, S.~Chu, L.~Camperlengo, R.~Anderson,
  Differentiating cause-of-death terminology for deaths coded as sudden infant
  death syndrome, accidental suffocation, and unknown cause: an investigation
  using {US} death certificates, 2003-2004, Journal of Forensic Sciences
  57~(364--369).

\bibitem{CDC-c}
{{Centers for Disease Control and Prevention}}, {{Injury Center - Data and
  Statistics(WISQARS) - WISQARS Data Visualization}},
  \url{https://www.cdc.gov/injury/wisqars/fatal.html}, Last accessed on
  2019-05-01 (2019).

\bibitem{CDC-d}
{{Centers for Disease Control and Prevention}}, {{Injury Center - Data and
  Statistics(WISQARS) - Nonfatal Injury 2000-2017}},
  \url{https://www.cdc.gov/injury/wisqars/fatal.html}, Last accessed on
  2019-05-01 (2019).

\end{thebibliography}

\end{document}